\documentclass[showpacs,showkeys,preprint]{revtex4}
\usepackage{graphicx}
\usepackage{amsmath}
\begin{document}
\title{Ripples in epitaxial graphene}
\author{F.Varchon}
\affiliation{Institut N\'eel, CNRS-UJF, BP 166, 38042 Grenoble Cedex 9 France\\}

\author{P.Mallet}
\affiliation{Institut N\'eel, CNRS-UJF, BP 166, 38042 Grenoble Cedex 9 France\\}

\author{J.-Y.Veuillen}
\affiliation{Institut N\'eel, CNRS-UJF, BP 166, 38042 Grenoble Cedex 9 France\\}

\author{L.Magaud}
\affiliation{Institut N\'eel, CNRS-UJF, BP 166, 38042 Grenoble Cedex 9 France\\}

\date{\today}

\begin{abstract}
Graphene outstanding properties directly come from its pecular electronic structure and thus from the honeycomb lattice symmetry. The way interaction with the substrate impact this lattice is of primary importance. This is peculiarly true for epitaxial graphene because of the SiC substrate. The advantage of this system that produces macroscopic samples in registry with a substrate could turn to a major drawback if the graphene lattice reveals to be strongly distorted. Extensive ab initio calculations supported by Scanning Tunneling Microscopy experiments, demonstrate here that the substrate indeed induces a strong nanostructuration of the interface carbon layer. It propagates to the above C layer where it generates incommensurate ripples in the honeycomb lattice.
\end{abstract}

\maketitle

\section{Introduction}

Between 2004 and 2005, three groups propelled graphene to the foreground ~\cite{Novoselov04,Novoselov,Zhang,Berger1}. They brought it from the theoretical to the real experimental world by synthetizing it for the first time. Indeed, since the pioneering work of Wallace in 1947 ~\cite{Wallace} very exotic properties had been forcast for this material that remained  untested. 3 years after its first synthesis, graphene is up to its promises : room temperature quantum Hall effect has been evidenced ~\cite{roomQHE} together with large coherence length and high electronic mobility ~\cite{Novoselov,Zhang,Berger_science}. Furthermore, these fundamental properties give a high application potential to graphene, especially in nanoelectronics (qbits, transistors) ~\cite{Novoselov_rise}. Outstanding electronic properties ~\cite{AWL,Backscatt,Wu} together with a high potential for application explain why graphene is becoming the new star of condensed matter physics.

Graphene is the name given to one isolated plane of carbon atoms arranged on a honeycomb lattice. The unit cell contains two equivalent carbon atoms (named A and B). The interaction with this pecular lattice makes electrons behave like relativistic massless fermions governed by a Dirac like equation. Indeed the band structure of graphene shows two bands with a linear dispersion that cross at the Fermi level. The crossing point is called the Dirac point. The anomalous quantum Hall effect of graphene for example is a direct consequence of this band structure.    

Two ways are used to synthetize high crystalline quality samples : either from mechanical exfoliation of highly oriented pyrolitic graphite (HOPG) ~\cite{Novoselov,Zhang} or from sublimation of Si from 4H or 6H SiC surface ~\cite{Berger1,Forbeaux,VanBommel}. We focus here on epitaxial graphene on SiC since then graphene is directly grown on a substrate and this approach is the only one that could lead to mass production for nanoelectronics needs. It has been shown to produce from few layers of graphite (FLG) to  one unique graphene layer ~\cite{Berger1,STMnous_PRB,Stroscio_science,Ohta_PRL}.

High quality graphene is required for transport properties and the development of nanoelectronic applications ~\cite{Novoselov_rise}. The mastering of the interface morphology is then a priority. The main question one has to answer is how the SiC substrate impacts the graphene unique structure. Indeed, Angle Resolved Photoelectron Spectroscopy (ARPES)  ~\cite{Ohta,Seyller,Bostwick,Emtsev,Ohta_PRL,Lanzara_gap} and transport measurements ~\cite{Berger_science} showed that the system morphology has a strong effect on the electronic structure of the FLG. The crystalline structure of the system has been approached by different techniques (x-ray diffraction ~\cite{Hass_APL,xray2,Charrier}, Low Energy Electron Diffraction (LEED) ~\cite{Chen,Rollings,Owman}, core level shift (CLS) ~\cite{Emtsev,Chen,Seyller, Rollings} and scanning tunneling microscopy (STM)  ~\cite{Chen,Tsai,Owman}). The existence of a $6\sqrt{3}\!\times\!6\sqrt{3}R30$ (hereafter 6R3) cell common to the SiC substrate and the graphene is rather well established for the Si-terminated surface. The C-terminated case is not so clear but the interface geometry does not seem to involve the 6R3 cell and calculations with this geometry are given here only for direct comparison.

Ab initio calculations using a highly simplified geometry for the interface (a $\sqrt{3}\!\times\!\sqrt{3}R30$ cell, hereafter R3) have established the bufferlayer role of the first C layer that electronically decouples the graphene planes from the substrate ~\cite{Varchon_PRL,Mattausch}. Here we address the actual $6\sqrt{3}\!\times\!6\sqrt{3}R30$ cell and on the basis of extensive ab initio calculations supported by scanning tunneling experiments we demonstrate that the substrate has a strong effect on the first two carbon layers. The interface C buffer layer is strongly distorted so that it cannot be seen as a graphene layer but it should instead be considered as the very beginnings of graphene. Its actual structure is elucidated at the atomic scale. Moreover, ripples are shown to be generated in the otherwise graphene like second plane.
 
Calculation and experimental details are given in part II. The third part presents the ab initio results for both Si- and C-terminated faces, for the buffer layer and for the first actual graphene layer. STM results are compared to ab initio calculations in part IV.

\section{Calculation and experimental details}

Ab initio calculations have been performed whithin the Density Functionnal Theory using the code VASP ~\cite{VASP}. Ultra soft pseudopotentials (USPP) ~\cite{USPP} have been used with a plane wave basis cutoff equal to 211 eV; The USPP have been extensively tested, especially, the C-short USPP was shown to correctly reproduce the band structure of graphene, graphite and bulk and surfaces of different SiC polytypes ~\cite{Varchon_PRL}. The Perdew and Wang ~\cite{PW} formulation of the General Gradient Approximation is used. Brillouin zone integration is performed using the $\Gamma$ point. 

Calculations have been performed with one or two carbon layers on the Si-terminated and C-terminated surfaces of SiC. The graphene and the (0001) SiC lattices are nearly commensurate with a 6R3 common cell with respect to SiC or a 13x13 cell with respect to graphene. In the following, periodicity expressed with respect to SiC (graphene) will be referred to -SiC (-G) for clarity. The calculations are performed in the actual 6R3-SiC interface geometry. We used complete and flat graphene layers as a starting point for all the C planes above the SiC substrate. This is to account for the ARPES results that demonstrated the existence of the $\sigma$ bands skeleton for the interface C layer ~\cite{Emtsev}: these bands are related to the presence of a well developped honeycomb like lattice.  When an additional graphene layer is added, the converged geometry of the interface C layer remains unchanged and forms the so-called bufferlayer.

The starting buffer layer system (1 C layer on SiC) was described by a supercell formed of 4 SiC bilayers plus one layer of C atoms arranged on a honeycomb lattice. H atoms are bound to the second SiC surface to passivate it and the whole cell contains 1310 atoms. The atomic positions where relaxed so that residual forces on the atoms are lower than 0.015 eV/\AA. The same cell with an additional graphene layer would be too large to keep the computational time within a reasonnable value (several months had already been necessary to reach convergence for each system) and we had to restrict the cell to two SiC bilayers, the buffer layer and the graphene layer (1216 atoms). Atoms in the lower SiC bilayer and the subsequent C plane are kept fixed to bulk positions. Because of the limited number of planes used to describe the substrate, forces could not be zeroed in these 3 layers that remain constrained. However, forces in the above graphene, buffer layer and last SiC plane could be relaxed to negligible values and the calculation that started from two flat honeycomb lattices reproduces the buffer layer geometry found with the first cell.

STM experiments were performed at room temperature, in ultra-high vacuum, using a home made microscope. Graphitization of n-type (nitrogen 1x$10^{18}$ $cm^{-3}$) 6H-SiC(0001) substrates was achieved by successive annealing into UHV, controlled by LEED and Auger spectroscopy. Experiments were done first on the buffer layer surface, and second on a mixed surface with mono- and bilayer graphene areas. The details of the growth process and the determination of layer thickness have been discussed elsewhere ~\cite{STMnous_PRB}. We used the symmetric STM atomic contrast to identify areas corresponding to monolayer graphene. Results have been checked on 2 different substrates with different tips and at different temperatures.

\section{Ab initio results}
\subsection{Buffer layer on the Si-terminated SiC surface}
 
Fig.1a shows a large scale image of the ab initio total charge density of an uncovered buffer layer (the picture is identical when it is covered by a graphene layer). It has three main characteristics:
  i) an obvious apparent 6x6-SiC modulation
  ii) The  6R3-SiC periodicity
  iii)low regions, separated by boundaries, that form nanograins with an unexpected local 2x2-G symmetry. This gives a mosaic structure to the buffer layer.

The 6x6-SiC modulation is strongly apparent (cell in red in Fig.1a) because it is related to the bright spots of Fig.1a and Fig.1b. It is not a real reconstruction since it is not compatible with translationnal symmetry within the buffer layer: Fig.1a and Fig.1b show that the local environment of the atoms at the 6x6-SiC bright spots is not the same. 
The 6R3-SiC periodicity (cell in blue in Fig.1a) corresponds to the SiC and graphene common cell. It is imposed by the calculation.

The mosaic pattern is composed of grains of more or less irregular hexagonal shape, 20\AA wide. The origin of the mosaic structure comes from the superposition of the C honeycomb and SiC lattices ~\cite{Chen}.  The two lattices do not adjust to each other. The grains (dark area) corresponds to regions where the SiC and honeycomb lattices match (in a local 2x2-G or R3-SiC symmetry). Si-C bonds are formed there. It generates the small hexagonal patterns of Fig.1a.  The local 2x2-G patterns are shifted in adjacent grains, the boundaries also accomodate for this shift. Atoms that are not in registry lay higher above the substrate and form boundaries (light area).

The Fourier transform of the charge density map of the free buffer layer (Fig.1d) and of the buffer layer covered with a graphene layer (not shown) are identical. Apart from the 1x1-G graphene and the 6x6-SiC spots, one remarkable feature of this FT image is the rather intense spots located around midway from the 1x1-G spots (but off-axis). These spots are also found in the FT of the STM images as shown below. They belong to the reciprocal lattice of the 6R3-SiC. Their intensity is locally enhanced (in k-space) by the presence of small grains with local 2x2-G (or R3 -SiC) symmetry. We call them 2x2-G spots in the following. Additionally, one remarks that the first order spots of the 6R3 reconstruction are missing at the centre of Fig.1d (although higher order spots of the 6R3 are present) ~\cite{noteTF}.

The corrugation of the buffer layer is rather large with a lower to upper atom height difference close to 0.12 nm (Fig.1f). In contrast to previous models proposed for the buffer layer ~\cite{Chen,Tsai}, the present structure  directly comes from atomic relaxation during ab initio calculation starting from a flat carbon honeycomb lattice on top of the SiC surface in the actual 6R3-SiC common cell. The resulting morphology is in agreement with CLS data that evidence the existence of 2 types of C atoms that are bound or not to Si atoms ~\cite{Emtsev,Chen}.

Ab initio calculations in the 2x2-G structure ~\cite{Varchon_PRL} demonstrated that the interface C layer has not the electronic structure of graphene but acts as a bufferlayer since it allows growth of subsequent C planes with graphene like dispersion . The present calculations, performed in the actual interface geometry, support this results. It demonstrates that part of the interface layer indeed has the simple 2x2-G geometry. It also further evidences the buffer role of the first C layer since a second C layer, on top of it, presents a (wavy) honeycomb lattice. 

\subsection{Graphene layer on the Si-terminated SiC surface}

The nanostructuration induced by the substrate in the bufferlayer propagates to the ontop graphene layer. It generates ripples that show up as a 6x6 modulation of the honeycomb lattice in the ab initio total charge density (Fig.2a, b and c).  This is confirmed by the Fourier transform that exhibits spots related to 1x1-G and 6x6-SiC periodicity (Fig.2d). High regions of the buffer layer correspond to high regions of the graphene layer (Fig.2c).The ripples have an amplitude of 0.04 nm for a wavelength of 1.9 nm. Such a long wavelength modulation does not discriminate between the two basis atoms, A and B, of the honeycomb lattice. So that no A versus B contrast can be seen on the ab initio total charge density map in agreement with STM image (inset Fig.4b).

An ab initio calculation of a free rippled graphene layer, where the atomic positions are frozen in the presently calculated 6x6 positions, reveals an ideal graphene like linear dispersion. The induced distorsion differenciates areas of the graphene layer but it does not differenciate on the whole A atoms from B atoms. The band structure calculation of the graphene layer with the substrate and buffer layer is unfortunately out of reach for the moment (more SiC bilayers need to be considered and the system would be too large for reasonable computational time). However, if a gap is opened in the graphene band structure because of the substrate, the phenomenon involved is more complex than a simple break of AB sublattice symmetry as it has been proposed recently ~\cite{Lanzara_gap}. Furthermore, ARPES ~\cite{Emtsev,Ohta_PRL}, STM ~\cite{Chen, STMnous_PRB} and preliminary calculations ~\cite{Varchon_PRL, Mattausch} demonstrate that the buffer layer has no $\pi$ states in the vicinity of the Fermi level. So these states cannot be involved in the disymmetrization of the A and B sublattices even if the buffer and the graphene layer stacking is graphite
 like (Bernal stacking).

\subsection{The C-terminated case}

The total charge density maps for the C-terminated interface are given for comparison with the Si-terminated interface. They give indications on what could happen at the C-terminated interface if the interface geometries were identical. In fact, the actual C-terminated geometry is not precisely known. The supercell used for these calculations is equivalent to the one used for the Si-terminated interface. The same complex mosaic structure appears for the first carbon layer (Fig.3a) but the hexagonal shape of the grains seems more regular and more pronounced. All boundaries intersections give rise to bright spots so that the 6x6 modulation is not so apparent. By the way, the nanostructuration of the graphene layer also appears to be stronger (Fig.3b). The difference between the two faces might come from the strength of the substrate - C layer bonds. In the C-terminated case, covalent C-C bonds are formed. They are stronger and shorter than Si-C bonds. The induced buffer layer lattice distorsions and subsequently, the graphene layer structuration are stronger.

\section{STM results on the Si-terminated surface}

The pecularities of the interface evidenced by ab initio calculations are also found in Constant Current topographic STM images of the buffer layer (Fig.4a), of the first graphene monolayer (Fig.4b) and of their respective FT (Fig.4c and 4d). The fact that STM images are related to the surface local density of states whereas total charge densities are obtained from ab initio calculations only has a minor effect in the present discussion. For the uncovered buffer layer, (the so-called nanomesh or 6R3 phase in the litterature ~\cite{Chen,Owman}), a honeycomb pattern with the 6x6-SiC periodicity is experimentaly observed (Fig.4a), with hexagons of irregular shapes in accordance with the calculation (Fig.1a). Although atomic resolution is hardly achieved on this surface ~\cite{STMnous_PRB,Chen,Owman}, the FT agrees with the 2x2-G periodicity calculated for the grains. Indeed, we find some tiny spots (circled in Fig.4c) at positions close to half the reciprocal vector of the 1x1-G lattice (k=29.5 $nm^{-1}$). The same spots are found on the FT of the corresponding calculated total charge shown in Fig.1d.

STM images of the first graphene layer capping the buffer layer strongly support our calculated results. A survey of the image of Fig. 4b, recorded at sample bias -0.2V,  reveals both the graphene 1x1-G honeycomb pattern superimposed to a 6x6-SiC superstructure. These features are recovered in the FT shown in Fig. 4d. A careful study of STM images of a few nm wide, like the 2.5 x 2.5 $nm^{2}$ image shown in inset, agrees with previous results of ~\cite{STMnous_PRB}: we do not find any detectable asymmetry between the adjacent carbon atoms of the graphene layer, and we use the symmetric contrast as a signature of the monolayer (an asymmetric contrast is found for graphene bilayer in AB stacking). As discussed in section IIIB, we believe that this is in contradiction with the mechanism suggested by Zhou et al  ~\cite{Lanzara_gap} for a possible gap opening at the Dirac point.

We can take advantage of the STM sensitivity to the interface states ~\cite{Charrier, Rutter2, Brar} to get new insights into the buffer layer structure capped by the graphene top-layer. At low sample bias (here -0.2V), the tip probes the $\pi$-like states of graphene, together with electronic states lying within the buffer layer or at the interface with SiC ~\cite{STMnous_PRB}. In the present case, interface states give rise to additional features in the FT of the STM image related to the buffer layer. As shown in Fig. 3d, we find a clear signature of the local 2x2-G symmetry, arising as pronounced bright spots (circle) close to the 2x2-G expected value. They correspond to the spots circled in the FT of the total charge density of the free buffer layer (Fig.1d). Both theory and STM experiment show that the buffer layer geometry is essentially unchanged when it is buried below one graphene monolayer.

From this results, it appears that the C buffer layer is not an intermediate sublimation phase but remains at the interface. A possible scenario for growth might proceed as follows: Si atoms sublimates, the C atoms rearrange and form the buffer layer. More Si atoms disappear, the C atoms freed form a new buffer layer while the former one flattens to form a graphene plane. Since the buffer layer needs to form prior to a new graphene layer appearance, a layer by layer growth is favoured as observed experimentally ~\cite{Charrier,Ohta_PRL}.

\section{conclusion}

As a conclusion, the C buffer layer is not an intermediate sublimation phase. It is always present at the Si-terminated SiC - graphene interface and it decouples the graphene layer from the substrate. Most of the system interest relies on the presence of this layer. It has a mosaic structure that derives from a honeycomb lattice distorted because of C-Si bond formation. Its structure is complex, it shows an 6x6-SiC modulation also found in STM images while the actual common cell corresponds to a $6\sqrt{3}\!\times\!6\sqrt{3}R30$-SiC. It also involves a local 2x2 symmetry (with respect to graphene) in the region where the honeycomb and the SiC lattices nearly match. This unexpected local symmetry appears in the Fourier transforms of the STM images confirming the existence of matching zones. The existence of this buffer layer agrees with a layer by layer growth mechanism where a new buffer layer is formed at the interface while the former one evolves into a graphene layer. The substrate induced deformation propagates to the first graphene layer where it creates incommensurate 6x6-SiC ripples. This incommensurate modulation, observed on STM images too does not break the AB symmetry. The effect of the ripples on transport properties requires further investigation. The possible use of the graphene and buffer layer substrate induced nanostructuration to form arrays of nanometric size islands also deserves further consideration.

{\bf Acknowledgement}

The authors acknowledge C.Naud, D.Mayou, P.Darancet, V.Olevano, E.Conrad, C.Berger and W.de Heer for fruitfull discussions. The present work was supported by the ANR GRAPHSIC project, the CIBLE07 project and computer grants at IDRIS-CNRS and CIMENT.

{\bf Figure captions}

Figure 1:(color on line) Buffer layer ontop of the Si-terminated SiC surface. The SiC - graphene common cell is shown in blue and the incommensurate 6x6-SiC modulation in red.
a) Mosaic structure of the  C buffer layer evidenced in this 11x11 $nm^2$ image of the total ab initio charge density, 0.04 nm above the upper atom.
b) total charge density in the 6R3-SiC unit cell.
c) cross section of the total charge density along the line define in b).
The black dots that appear when the cross section goes through the middle of an atom are due to the use of pseudopotentials (no core electrons).
d) Fourier transform of the large scale image of the total charge density a). The solid line arrow points to a 1x1-G spot, the dashed line arrow to a 6x6-SiC spot and the spots in the circle corresponds to the local 2x2-G periodicity. 
e) positions of the atoms in the unit cell, Si atoms in the last SiC plane are in blue, the colour of the atoms in the buffer layer varies as a fonction of their height, ranging from green (close the substrate) to  red (uppermost atom).
f) Height profile of the buffer layer atoms along the line defined in e. The line is just a guide for the eyes.

Figure 2: (color online) Graphene layer above the C buffer layer for the Si-terminated surface. a) large scale image (11x11 $nm^2$) of the total charge density 0.04 nm above the uppermost atom in logarithmic scale. It exhibits the characteristic 6x6-SiC periodicity (cell shown in red). The 6R3-SiC common cell is shown in blue.
b) total charge density in the 6R3-SiC unit cell.
c) cross section of the total charge density along the line define in b). The section is not taken at the same position than in Fig.1c to focus on the graphene layer.
d) Fourier transform of the large scale image of the total charge density a). The solid line arrow points to a 1x1-G spot, the dashed line arrow to a 6x6-SiC spot. 
e) positions of the atoms in the unit cell, C atoms in the buffer layer are in blue, the colour of the atoms in the graphene layer varies as a fonction of their height, ranging from  green (close the buffer layer) to  red (uppermost atom).
f) Height profile of the graphene layer atoms along the line define in e.

Figure 3:(color online) Morphology of the buffer layer and the ontop graphene layer for the C-terminated SiC surface.a) charge density map of the interface buffer layer (8x8 $nm^2$), b) charge density map of the first ontop graphene layer (8x8 $nm^2$).

Figure 4: (color online) STM images and their related Fourier transforms of the buffer layer and the graphene layer on the Si-terminated surface. a) 12x12 $nm^2$ large STM image of the buffer layer (V=-2.0 V and I=0.5 nA). It clearly shows the irregular hexagonal pattern of the 6x6 -SiC periodicity.
b) 12x12 $nm^2$ STM image of the graphene monolayer (V=-0.2 V, I=0.1 nA) (inset 2.5x2.5 $nm^2$). 
c) Fourier transform of image a), the dashed line arrow points to a 6x6-SiC spot and the 2x2-G spots are circled.
d) Fourier transform of image b), the arrow points to the 1x1-G spot, the dashed arrow points to the 6x6-SiC spots and 2x2-G related spots are circled.
\newpage
\begin{figure}
\includegraphics[width=18.cm]{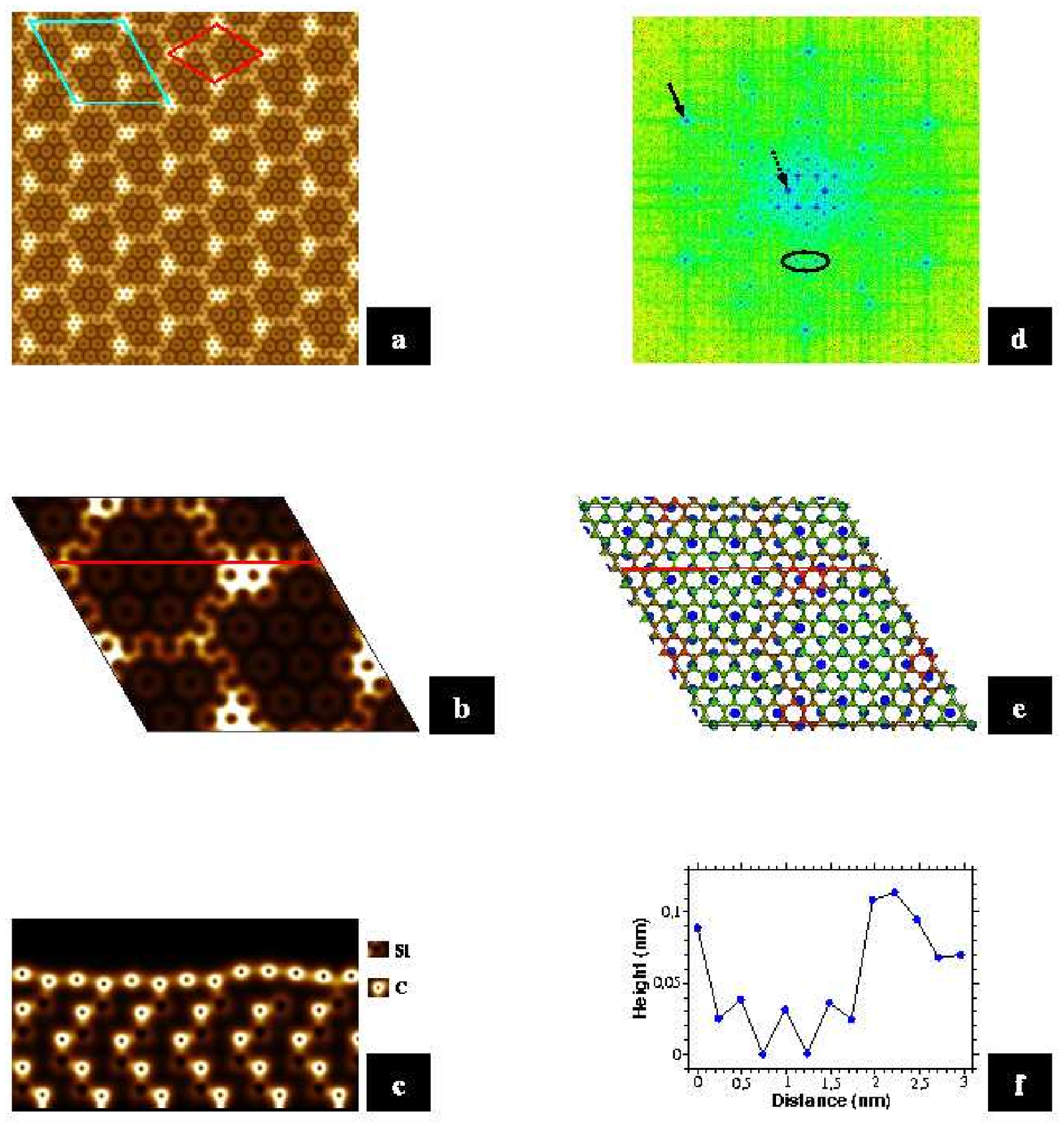}
\label{fig:fig1}
\end{figure}

\begin{figure}
\includegraphics[width=18.cm]{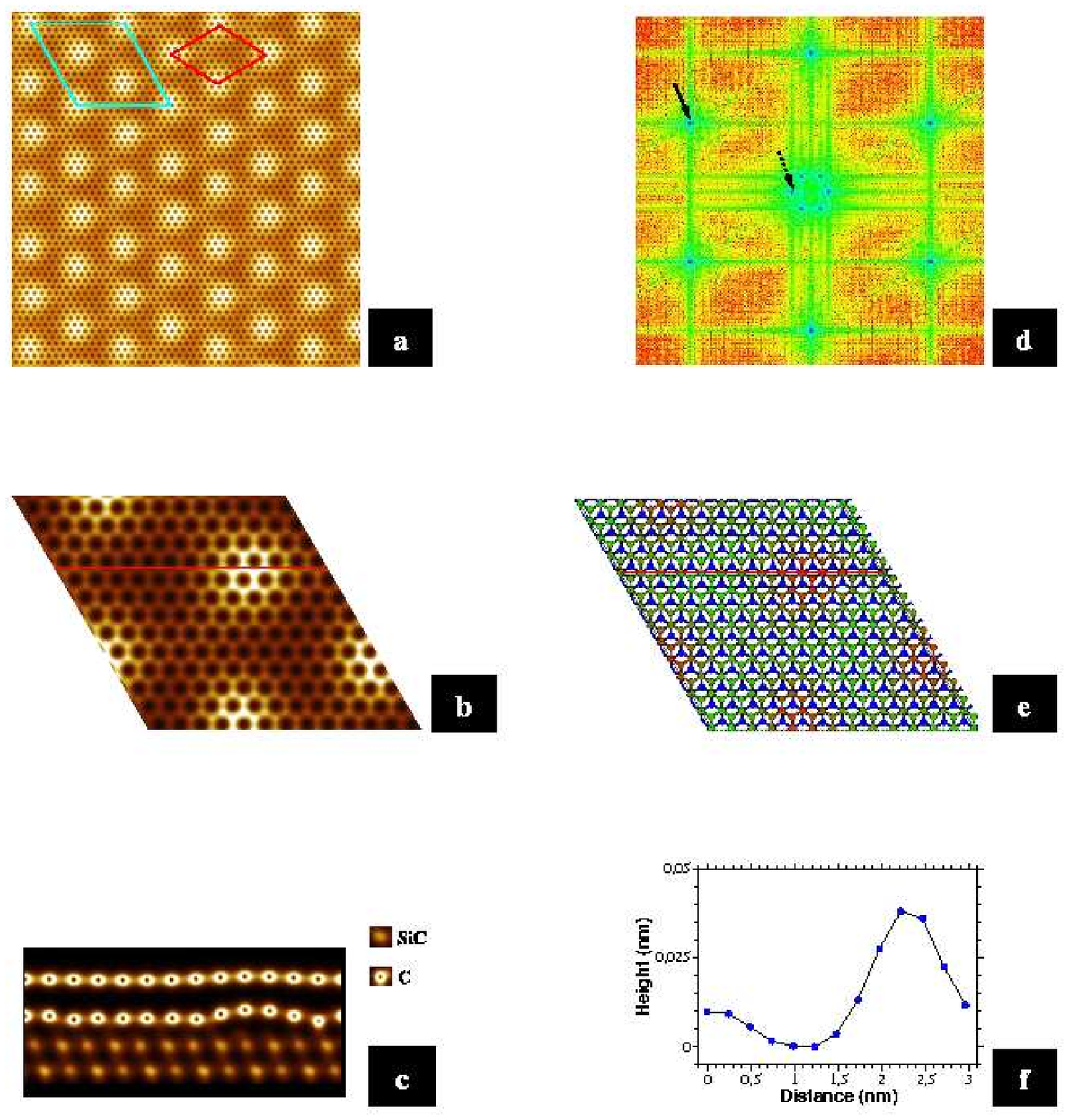}
\label{fig:fig2}
\end{figure}

\begin{figure}
\includegraphics[width=18.cm]{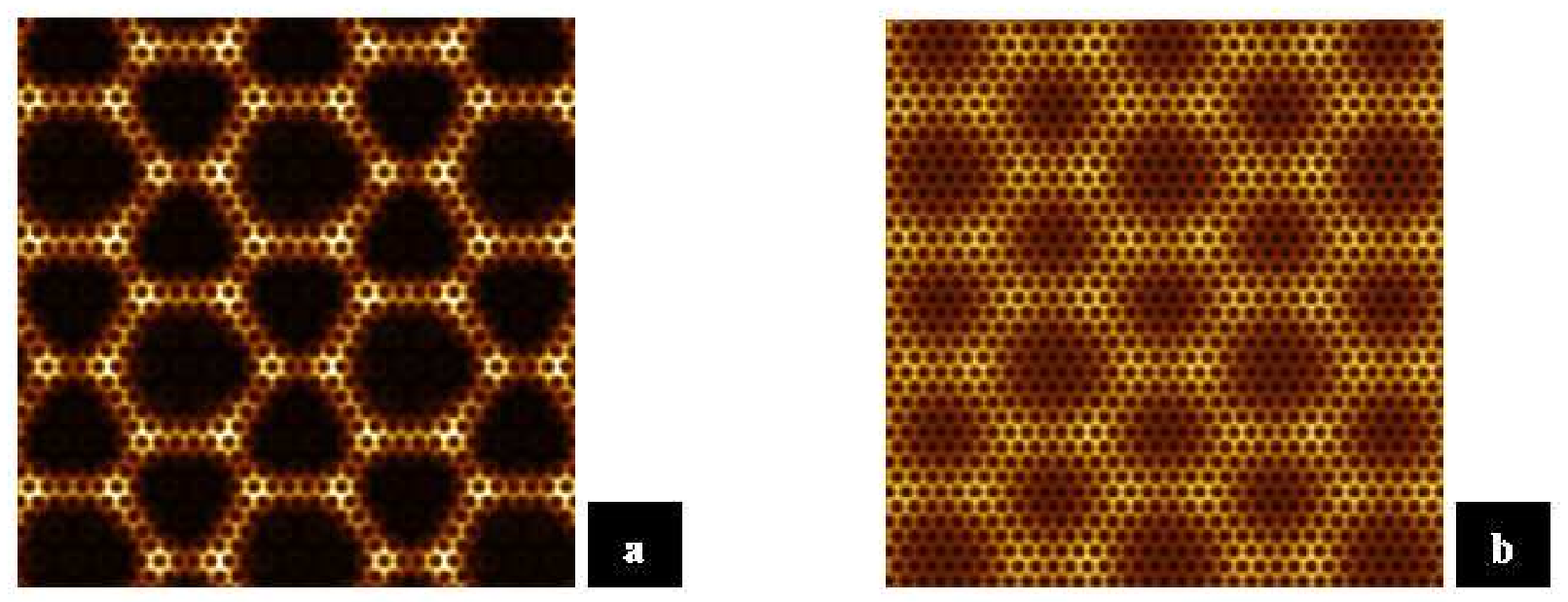}
\label{fig:fig3}
\end{figure}

\begin{figure}
\includegraphics[width=18.cm]{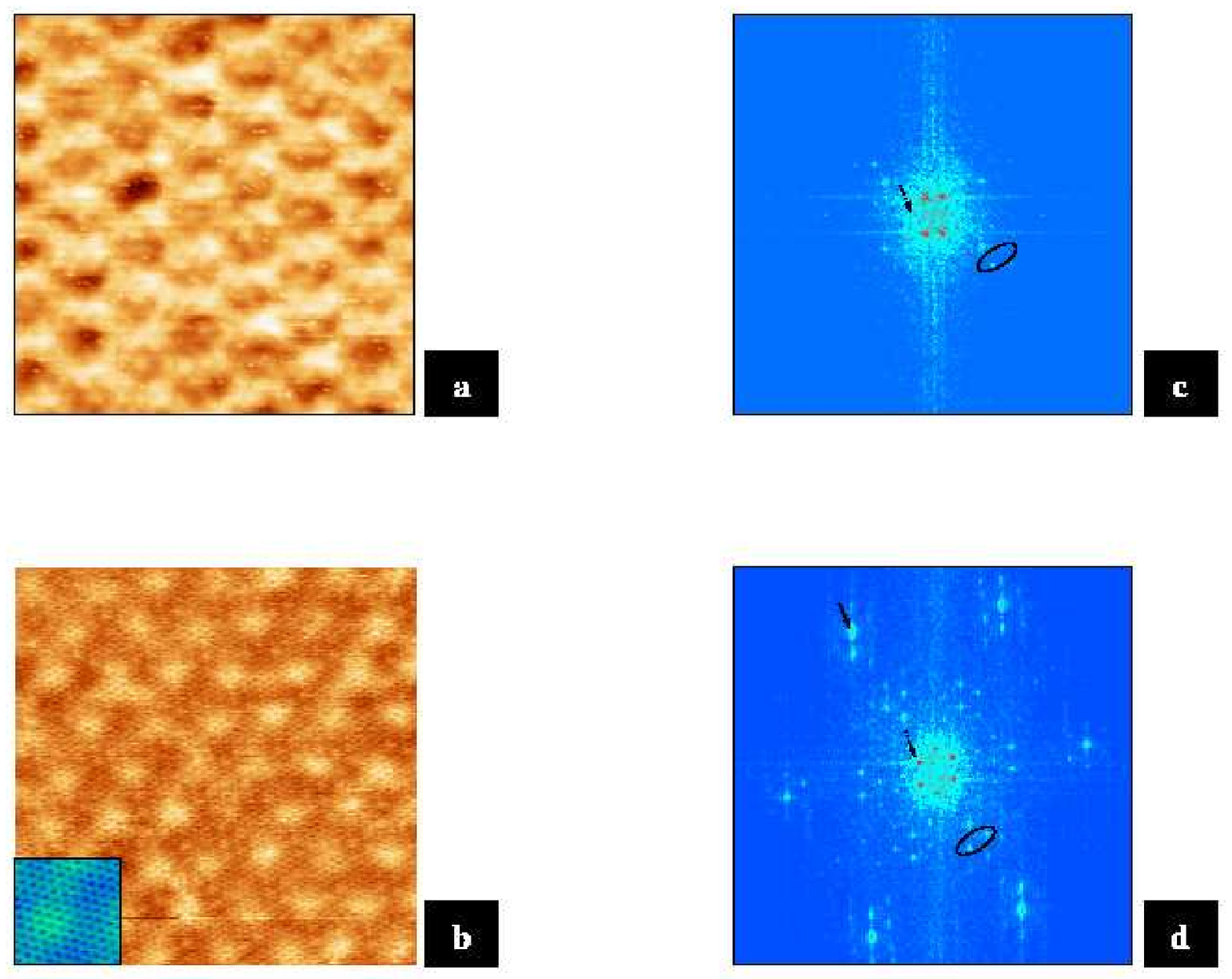}
\label{fig:fig4}
\end{figure}


\begin{thebibliography}{99}

\bibitem{Novoselov04}
K.~S.Novoselov {\it et al.} {\it Science} {\bf 306}, 666-669 (2004)

\bibitem{Novoselov}
K.~S.Novoselov {\it et al.} {\it Nature} {\bf 438}, 197-200 (2005)

\bibitem{Zhang}
Y.Zhang {\it et al.} {\it  Nature} {\bf 438}, 201-204 (2005)

\bibitem{Berger1}
C.Berger {\it et al.} {\it J.Phys. Chem.B} {\bf 108}, 19912-19916 (2004)

\bibitem{Wallace}
P.~R.Wallace  {\it Phys. Rev.} {\bf 71}, 622-634 (1947)

\bibitem{roomQHE}
K.~S.Novoselov {\it et al.} {\it Science} {\bf 315}, 137 (2007)

\bibitem{Berger_science}
C.Berger {\it et al.} {\it Science} {\bf 312}, 1191-1196 (2006)

\bibitem{Novoselov_rise}
A.~K.Geim  and K.~S.Novoselov, {\it  Nature Mat.} {\bf 6},183-191 (2007)

\bibitem{AWL}
S.~V.Morozov {\it et al.} {\it Phys.Rev.Lett.} {\bf 97},016801 (2006)

\bibitem{Backscatt}
T.Ando, T.Nakanishi, R.Saito {\it J.Phys.Soc.Jpn} {\bf 67}, 2857-2862 (1998)

\bibitem{Wu}
X.Wu {\it et al.} {\it Phys. Rev. Lett.} {\bf 98}, 136801 (2007)

\bibitem{VanBommel}
A.~J.Van Bommel, J.E.Crombeen and A. van Tooren, {\it Surf.Sci.}{\bf 48}, 463-472 (1975)

\bibitem{Forbeaux}
I.Forbeaux, J.~-M.Themlin, J.~-M.Debever, {\it  Phys. Rev. B} {\bf 58}, 16396-16406 (1998)

\bibitem{STMnous_PRB}
P.Mallet {\it et al.} {\it Phys. Rev. B} {\bf 75}, 205315 (2007)

\bibitem{Stroscio_science}
G.M.Rutter  {\it et al.} {\it Science} {\bf 317}, 219-222 (2007)

\bibitem{Ohta_PRL}
T.Ohta {\it et al.} {\it Phy. Rev. Lett.} {\bf 98}, 206802 (2007) 

\bibitem{Ohta}
T.Ohta {\it et al.} {\it Science} {\bf 313}, 951-954 (2006)

\bibitem{Seyller}
Th.Seyller {\it et al.} {\it Surf. Sci.} {\bf 600}, 3906-3911 (2006)

\bibitem{Bostwick}
A.Bostwick {\it al.} {\it Nature Phys.} {\bf 3} 36-40 (2006)

\bibitem{Emtsev}
E.~K.Emtsev {\it et al.} {\it Mater. Sci. Forum} {\bf 556-557}, 525 (2007)

\bibitem{Lanzara_gap}
S.Y.Zhou {\it et al.} {\it Nature Mat.} {\bf 6}, 771-776 (2007)

\bibitem{Hass_APL}
J.Hass {\it al.}  {\it Appl. Phys. Lett.} {\bf 89}, 143106 (2006)

\bibitem{xray2}
J.Hass {\it et al.} {\it Phys. Rev. B} {\bf 75}, 214109 (2007). 

\bibitem{Charrier}
A.Charrier {\it et al.} {\it J. Appl. Phys.} {\bf 92}, 2479-2484 (2002)

\bibitem{Chen}
W.Chen {\it et al.} {\it Surf. Sci.} {\bf 596}, 176-186 (2005)

\bibitem{Rollings}
E.Rollings {\it et al.}  {\it J.Phys.Chem.Sol.} {\bf 67}, 2172 (2006)

\bibitem{Owman}
F.Owman, P.Martensson, {\it Surf. Sci.} {\bf 369}, 126-136 (1996)

\bibitem{Tsai}
M.~-H.Tsai {\it et al.} {\it Phys. Rev. B} {\bf 45}, 1327-1332 (1992)

\bibitem{Varchon_PRL}
F.Varchon {\it et al.} {\it Phys. Rev.Lett.}{\bf 99}, 126805 (2007)

\bibitem{Mattausch}
A.Mattausch and O.Pankratov, {\it Phys.Rev.Lett.} {\bf 99}, 076802 (2007)

\bibitem{VASP}
G.~Kresse and J.~Hafner, {\it Phys. Rev. B} {\bf 47}, 558 (1993)

\bibitem{USPP}
G.~Kresse and J.~Hafner, {\it J.Phys. Condens. Matter} {\bf 6}, 8245 (1994)

\bibitem{PW}
J.~P.~Perdew and Y.~Wang, {\it Phys. Rev. B} {\bf33}, 8800 (1986)

\bibitem{noteTF}
This is not an artifact of the FT software, since when one selects only one family of 2x2-G patches arranged in a 6R3-SiC network in the image before performing the FT, the first order 6R3-SiC spots reappear. It is therefore an intrinsic property of the image, and we believe this might be related to the spatial arrangement of the 3 families of local 2x2-G grains. First order spots of the 6R3-SiC reconstruction are also absent in FT of the STM images. 

\bibitem{Rutter2}
G.M.Rutter et al.,  ArXiv:cond-mat/0711.2523

\bibitem{Brar}
V.W.Brar et al, {\it Appl.Phys.Lett} {\bf 91}, 122102 (2007)


\end{thebibliography}
\end{document}